# Variability of Disk Emission in Pre-main Sequence and Related Stars. V. Occultation Events from the Innermost Disk Region of the Herbig Ae Star HD 163296

Monika Pikhartova[1,2], Zachary C. Long[2], Korash D. Assani[3], Rachel B. Fernandes[4,5], Ammar Bayyari[6],
Michael L. Sitko[2,7], Carol A. Grady[8,9], John P. Wisniewski[10], Evan A. Rich[10,11], Arne A. Henden[12], and
William C. Danchi[9]

[1] Anton Pannekoek Institute for Astronomy, University of Amsterdam, Science Park 904, 1098 XH Amsterdam, The Netherlands; monika.pikhartova@gmail.com
[2] Department of Physics, University of Cincinnati, Cincinnati, OH 45221, USA
[3] Department of Astronomy, University of Virginia, Charlottesville, VA 22904, USA
[4] Lunar and Planetary Laboratory, The University of Arizona, Tucson, AZ 85721, USA
[5] Earths in Other Solar Systems Team, NASA Nexus for Exoplanet System Science, Tucson, AZ, USA
[6] Department of Physics and Astronomy, University of Hawaii, Honolulu, HI 96822, USA
[7] Center for Extrasolar Planetary Systems, Space Science Institute, 4750 Walnut Street, Suite 205, Boulder, CO 80301, USA
[8] Eureka Scientific, 2452 Delmer Street, Suite 100, Oakland, CA 96402, USA
[9] Exoplanets and Stellar Astrophysics Laboratory, Code 667, Goddard Space Flight Center, Greenbelt, MD 20771, USA
[10] Homer L. Dodge Department of Physics and Astronomy, University of Oklahoma, Norman, OK 73019, USA
[11] Department of Astronomy, University of Michigan, 1085 S. University, Ann Arbor, MI 48109, USA
[12] American Association of Variable Star Observers, 49 Bay State Road, Cambridge, MA 02138, USA
Received 2020 June 23; revised 2021 May 11; accepted 2021 May 19; published 2021 September 24

## Abstract

HD 163296 is a Herbig Ae star that underwent a dramatic ∼0.8 magnitude drop in brightness in the V photometric band in 2001 and a brightening in the near-IR in 2002. Because the star possesses Herbig–Haro objects traveling in outflowing bipolar jets, it was suggested that the drop in brightness was due to a clump of dust entrained in a disk wind, blocking the line of sight toward the star. In order to quantify this hypothesis, we investigated the brightness drop at visible wavelengths and the brightening at near-IR wavelengths of HD 163296 using the Monte Carlo Radiative Transfer Code, HOCHUNK3D. We created three models to understand the events. Model 1 describes the quiescent state of the system. Model 2 describes the change in structure that led to the drop in brightness in 2001. Model 3 describes the structure needed to produce the observed 2002 brightening of the near-IR wavelengths. Models 2 and 3 utilize a combination of a disk wind and central bipolar flow. By introducing a filled bipolar cavity in Models 2 and 3, we were able to successfully simulate a jet-like structure for the star with a disk wind and created the drop and subsequent increase in brightness of the system. On the other hand, when the bipolar cavity is not filled, Model 1 replicates the quiescent state of the system.

*Unified Astronomy Thesaurus concepts:* Protoplanetary disks (1300); Circumstellar disks (235)

## 1. Introduction

HD 163296 is a Herbig Ae star that has come under intense scrutiny in the past decade due to changes in its brightness, small-scale structure, disk illumination, and bipolar jet outflow. More recently, imaging of the disk, its rings, and possible planets have been studied in abundance. Near-IR (NIR) observations revealed a 64 au scale inner dust ring (Garufi et al. 2017; Monnier et al. 2017; Muro-Arena et al. 2018), 1.3 mm continuum ALMA imaging revealed three azimuthal gaps in the disk (Isella et al. 2016). The second generation Near Infrared Camera of the W.M. Keck Observatory (Keck/NIRC2) and the Atacama Large Millimeter/submillimeter Array (ALMA) data led to the discovery of a protoplanet candidate outside the inner ring and suggestions of Jovian planets on wider orbits (Guidi et al. 2018; Pinte et al. 2018; Teague et al. 2018). Unlike the aforementioned works, this paper concentrates on the innermost region of the disk and its variations.

In 2002, there was an ∼50% increase in its brightness at NIR wavelengths (Sitko et al. 2008). Around the same time, Tannirkulam et al. (2008), using data obtained in 2001 using the Keck Interferometer (Colavita et al. 2013), found an ∼20% increase in the derived size of the structure in the inner region. Wisniewski et al. (2008) found evidence for possible changes of illumination of its circumstellar disk as seen in scattered light. Furthermore, Rich et al. (2019) later indicated the presence of time variable, nonazimuthally symmetric illumination of the outer disk. Such changes are mostly due to dust casting shadows from the inner region of the disk. Possible mechanisms include "puffed up" inner disk rim, clumpy dusty disk winds or warped inner disks (e.g., see Natta et al. 2001; Dullemond & Dominik 2004a; Debes et al. 2017; Benisty et al. 2018; Casassus et al. 2018).

Wassell et al. (2006) noted that the object possesses bipolar outflows with condensations (Herbig–Haro objects) suggesting the possibility of episodic ejections of material. Ellerbroek et al. (2014) measured the proper motions of these condensations and suggested that they could be traced to specific epochs. One recent "ejection" occurred at about the same time as a 0.8 magnitude drop in brightness in the V photometric band, recorded by the All Sky Automated Survey (ASAS; Pojmanski & Maciejewski 2004), between 2001 and 2002. They suggested that the drop in brightness might have been due to dust entrained in a variable disk wind, not unlike the general dust+gas disk wind model described by Bans & Königl (2012), as can be seen in Figure 1. This general model is supported by the detection of a rotating molecular disk wind using ALMA (Klaassen et al. 2013).

One aspect of the earlier works on the observed photometric and interferometric changes (Sitko et al. 2008; Tannirkulam et al. 2008) that was not recognized was that the Keck Interferometry observations were mostly sensitive to structure along the minor axis of the disk, along the direction of the bipolar jets in HD 163296, which is 90° from the position angle of the disk. Because of this, and the need for significant attenuation of the light from the star, we investigated the effect that a jet-like structure similar to that proposed by Ellerbroek et al. (2014) would have on both the visible light as well as the thermal emission of the disk.

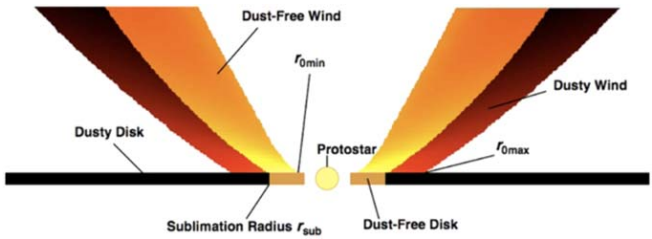

**Figure 1.** The Bans and Königl general disk wind model for the NIR excess emission in protostars used as a base for our model. The schematic shows a centrifugally driven disk wind from the vicinity of a low-luminosity protostar. Dust is being lifted from the disk surface beyond the sublimation radius. In the innermost region, the wind is gaseous, and further on it becomes dusty. Bans & Königl (2012) state that the outer radius might be limited to ∼2 sublimation radii because of the dust absorption of stellar FUV photons in the wind. Dusty wind reprocessing stellar radiation produces most of the NIR excess emission.

In this paper, we investigate whether all of these phenomena might be related. Specifically, we model the disk and jet using a Monte Carlo radiative transfer model that is capable of producing the drop in brightness as well as changes in the innermost regions of the disk+jet, and would lead to a NIR brightness change. As HD 163296 has undergone additional photometric changes after 2002, and may continue to do so in the near future, the model makes predictions that can be compared with new observations. We produce three separate models for HD 163296: Model 1 for its quiescent (2005) state, Model 2 for the 2001 drop in brightness in V band, and Model 3 for the subsequent brightening of the NIR region observed in 2002. We compare them to the spectral energy distributions where available.

## 2. Observations

Visible through NIR spectra of HD 163296 were obtained on 10 nights between 2008 and 2016 using the SpeX spectrograph (Rayner et al. 2003) on NASA's Infrared Telescope Facility (IRTF). The SpeX observations were made using the cross-dispersed echelle gratings between 0.8 and 5.4 μm, using a spectrograph slit width of 0.8″. Telluric corrections and flux calibration were produced using the Spextool reduction package (Vacca et al. 2003; Cushing et al. 2004) running under IDL. For HD 163296, the A0V star HD 163336 was used as reference. The observing and data reduction procedures were the same as previously described by Sitko et al. (2012), as was the technique to produce the absolute flux calibration. The dates (UT) and airmass of both HD 163296 and HD 163336 are listed in Table 4 in Appendix A. From these data, we were also able to determine the mass accretion rate, as described in Appendix D.

Additionally, we included the spectra obtained in 2002 and 2005, using the Lick Observatory 3 m telescope. In 2002, the Near Infrared Imaging Spectrograph (NIRIS) was used, while in 2005, its upgrade, the Visible and Near-Infrared Imaging Spectrograph (VNIRIS) was used. The observations are described further in Sitko et al. (2008).

We also obtained spectrophotometry between 3 and 13 μm using the Aerospace Corporation's Broad-band Array Spectrograph

**Table 1**
Modeling Stellar Parameters

| Parameter | Value | Reference |
|---|---|---|
| Age | $9.0 \pm 0.5$ Myr | (1) |
| Distance | 101.5 pc | (1) |
| Mass | 1.71 $M_\odot$ | (1) |
| Radius | 2.14 $R_\odot$ | (1) |
| Luminosity | $18.6 \pm 0.4$ $L_\odot$ | (1) |
| Temperature | 9500 K | (2) |
| B | 8.09 | (3) |
| B-V | 0.08 | (3) |
| Disk Inclination | 45° | (4) |

**Note.** References: (1) Either directly taken from Gaia Collaboration et al. (2018) or calculated using Gaia DR2 Data; (2) Guimarães et al. (2006); (3) Ducati (2002); (4) Isella et al. (2007); Qi et al. (2011). Latest publications seem consistent with 46° (Kluska et al. 2020).

System (BASS). BASS uses a cold beam splitter to separate the light into two separate wavelength regimes. The short-wavelength beam includes light from 2.9 to 6 μm, while the long-wavelength beam covers 6–13.5 μm. Each beam is dispersed onto a 58-element Blocked Impurity Band (BIB) linear array, thus allowing for simultaneous coverage of the spectrum from 2.9 to 13.5 μm. The spectral resolution $R = \lambda/\Delta\lambda$ is wavelength-dependent, ranging from about 30 to 125 over each of the two wavelength regions (Hackwell et al. 1990).

In order to construct a complete spectral energy distribution for constraining models of the HD 163296 system, we also include a variety of archival data, both from the Simbad/Vizier SED tool and other sources in the literature—the NASA/IPAC Infrared Science Archive (Spitzer, WISE, MSX, 2MASS, IRAS, Akari IRC, Herschel PACS), the Mikulski Archive for Space Telescopes (IUE, Hubble/STIS), and millimeter observations from Isella et al. (2007); Henning et al. (1994, 1998); Mannings (1994); and Sandell et al. (2011).

## 3. Radiative Transfer Modeling

In order to model the SEDs of HD 163296 we used the three-dimensional Monte Carlo radiative transfer (MCRT) code, HOCHUNK3D, of Whitney et al. (2013). We adopted the distance of 101.5 pc (Gaia Collaboration et al. 2018). Using a spectral type of A1V, and intrinsic pre-main sequence values of the effective temperature, bolometric corrections from Pecaut & Mamajek (2013), and observed B-V colors from Ducati (2002), we adopted an effective temperature of 9500 K since it was the closest photospheric file from the 9400 K of Guimarães et al. (2006), and derived stellar radius 1.71 $R_\odot$. Using the pre-main sequence evolutionary tracks of Tognelli et al. (2011) we derived a stellar mass of 2.14 $M_\odot$ (see Appendix C). The limb darkening of the central source was included in the code. The stellar parameters are summarized in Table 1.

HOCHUNK3D allows for each of the two independent, constituent coplanar disks to have its own set of input parameters that dictate the structure and composition of that disk. These include different vertical scale heights to simulate grain growth and settling (Dullemond & Dominik 2004a, 2004b) as well as different radial extents. Both disks also include polycyclic aromatic hydrocarbons and very small grains (PAH/VSG) dust model from Draine & Li (2007), which is discussed in Wood et al. (2008).

**Table 2**
Modeling Disk Parameters

| Parameters | Values |
| --- | --- |
| Outer Edge of the Disk | 450.1 au |
| Small Grain Disk Density Scale Height | 0.4 $R_{sub}$ |
| Small Grain Disk Density Scale Height at 0.5 au | 0.5 au |
| Small Grain Disk Density Exponent | 1.88 |
| Small Grain Disk Density Scale Height Exponent | 1.045 |
| Small Grain Disk Minimum Radius | 49 au (150.1 $R_{sub}$) |
| Small Grain Disk Outer Radius | 450.1 au |
| Large Grain Disk Density Scale Height | 0.7 $R_{sub}$ |
| Large Grain Disk Density Scale Height at 0.5 au | 0.7 au |
| Large Grain Disk Radial Density Exponent | 0.1 |
| Large Grain Disk Density Scale Height Exponent | 0.45 |
| Large Grain Disk Minimum Radius | 1.3 au (4 $R_{sub}$) |
| Large Grain Disk Outer Radius | 450.1 au |
| Minimum Envelope Radius | 0.3 au (0.9 $R_{sub}$) |
| Outer Edge Envelope Radius | 450.1 au |
| Minimum Envelope Gap Radius | 1.05 au |
| Outer Edge Envelope Gap Radius | 450.1 au |
| Gap Density | 0 |

To guide our modeling, we used the combined ALMA and the Karl G. Jansky Very Large Array (VLA) data of Guidi et al. (2016) and the mid-IR VLT interferometry from van Boekel et al. (2004). For the ALMA continuum maps at 0.85 mm and 1.3 mm, the emission extends to at least 200 au, while the 9 mm emission is largely confined to a region about one-tenth of that size. The concentration of the longest-wavelength emission to the inner regions is consistent with the interferometrically resolved silicate band emission from the Very Large Telescope Interferometer (VLTI), where grains located closer to the star were larger than those found further out (van Boekel et al. 2004).

We assigned one disk to contain large grains with a scale height 0.7 sublimation radii ($R_{sub}$) in hydrostatic equilibrium (HSEQ), and the other disk to contain smaller grains with a scale height of 0.4 $R_{sub}$ in HSEQ, which overlaps the first disk. $R_{sub}$ is the typical radius where the silicate grains have a typical temperature of 1500 K. The disk scale height scales as approximately $H_{disk} \sim r^b$, while the radial density scales as $\rho_{disk} \sim r^{-a}$. These scale heights are approximately 0.5 au for the small grain disk and 0.7 au for the large grain disk at a distance of 0.5 au. The values used are shown in Table 2. The overall disk mass is 0.19 $M_\odot$, where 0.35 is the fraction of the disk mass of the large grain settled disk. The scale heights are defined at the inner edge of each disk. The small grain disk has an inner radius of 49 au (150.1 $R_{sub}$), while the large grain disk starts at 1.3 au (4 $R_{sub}$). The combination of the inner radii and disk shape ensures the small grains lie above the large grains in the outer disk. Code assumes the $R_{sub}$ for the small grains based on $T_{sub}$ that equals 1500 K. But larger grains can exist closer in than the small grain $R_{sub}$, since they reach $T_{sub}$ at radii smaller than $R_{sub}$, as shown by observations in van Boekel et al. (2004). Some of the grains may be super-refractory grains and the fact that larger grains allow them to survive closer to the star than smaller grains of the same composition (Benisty et al. 2010). More discussion on dust extinction can be found in Appendix B.

For the large dust grains, we used a mixture of amorphous carbon and astronomical silicates that have a power-law size distribution with exponents of 3.0, plus exponential cutoff with a turnover at 50 $\mu$m, and a maximum size of 1000 $\mu$m (Model 1 from Wood et al. 2002). The larger grains are used to represent the effects of grain growth and the settling of dust to the disk midplane. The disk radius was set up from 1.3 au (4 $R_{sub}$) to 450.1 au. For the small grain component we used the dust distribution of Kim et al. (1994), which are normal interstellar medium grains of graphite and silicates. The size distribution is roughly a power law with an exponent of 3.5. The grain sizes go from around 0.02 to 1 $\mu$m. The radius went from 150.1 $R_{sub}$ to 450.1 au.

Our model includes a third disk component that was originally developed for an envelope of material infalling onto the disk but that we employed to create an inner disk fan representing either a region with enhanced scale height or an inner disk dusty wind (for example, see Figure 1 in Bans & Königl 2012, and Figure 9 in Ellerbroek et al. 2014), to produce the occulting event seen in 2001. The envelope employs the same dust distribution of Kim et al. (1994) as the small grain disk and the opacity file as both coplanar disks. The envelope covers a radial range from 0.3 au (0.9 $R_{sub}$) to 1.05 au, with a gap extending to 450.1 au that has a constant scale density. A similar gap was used in Sitko et al. (2008) when it was needed for extending the envelope past the outer radius of the disk. The envelope is set up with a simple power-law formation geometry, $\rho_1 r^{-d}$, where $\rho_1$ is the fiducial density at 1.3 au (units of g cm$^{-3}$) and $d$ is the density exponent. How those variables change from one epoch to another is shown in Table 3. The ambient density, which is the floor for the envelope density, is set up to be $1 \times 10^{-23}$ g cm$^{-3}$.

When the envelope is combined with the bipolar cavity, we can model a nondisk jet or a disk wind, which is standard in Whitney et al. (2013) models. The task of the components was to make the disk wind be able to get in the line of sight between the observer and the star, by changing its height. The bipolar cavity employs both the Kim et al. (1994) dust distribution and the Draine & Li (2007) opacity dust file. The envelope is carved out as a polynomial-shaped cavity. The opening angle of both the inner and outer cavity surface changes based on the epoch we are modeling, as well as their cavity exponents that act as approximately $w^f$, where $w$ is the width of the cavity versus the vertical distance, and the $f$ coefficient stands for inner or outer cavity density distributions. Both the inner and outer cavity surface at $w = 0$ have a z-axis intercept at −0.13 au, which means it has a nonzero value in the disk midplane. The exponent for cavity density power law is set up as $\rho = xr^f$, where $x$ is the coefficient from the inner cavity density distribution.

The model does not account for any outside illumination, fractal clumping, disk accretion, gaps, or spiral warps in the disk, and does not solve for vertical hydrostatic equilibrium.

### 3.1. Model 1: Quiescent State Modeling

We first reproduced the spectral energy distribution of HD 163296 in its quiescent state. The adopted stellar parameters are shown in Table 1 and the disk parameters we adjusted to achieve the best visual agreement with observations are in Table 2. Both the stellar and the disk parameters were not altered between the quiescent state model (Model 1, 2005) and the other models (Model 2, 2001, and Model 3, 2002).

Since the changes in the SED were observed on such a short timescale (∼1 yr), the structural changes are almost surely occurring close to the star, such as the disk fan or the base of a jet. Beyond this radial distance, the Keplerian timescale would be too long for any variability to be observed. In reality, changing the base can affect other parts of the jet. Most of the effects will be visible from material that is hot, bright, and dense, i.e., the base, which is what we produce in the further models. The parameters that we varied between each model are shown in Table 3. Parameters that affect the entire disk were held constant, as they cannot change on short timescales.

Table 3
Variable Model Parameters

| Parameters | Quiescent Model | 2001 Model | 2002 Model |
|---|---|---|---|
| Density at 1.3 au | $9.5 \times 10^{-16}$ | $3.3 \times 10^{-16}$ | $3.3 \times 10^{-16}$ |
| Exponent of envelope power-law density | −14 | 2.8 | 2.8 |
| Opening angle of inner cavity surface | 15° | 5.5° | 14° |
| Cavity shape exponent inner surface | 1.25 | 1.41 | 1.31 |
| Opening angle of outer cavity surface | 15° | 1° | 1° |
| Cavity shape exponent outer surface | 1.25 | 2.1 | 2.1 |
| Exponent for cavity density power law | 1 | 1.5 | 1.5 |
| Coefficient for inner cavity density distribution | 0 | $1.5 \times 10^{-14}$ | $1.5 \times 10^{-14}$ |
| Coefficient for outer cavity density distribution | 0 | $1.5 \times 10^{-14}$ | $1.5 \times 10^{-14}$ |

In Figure 2 we present our SED model of the quiescent state of the star. The shorter wavelengths do not fit because the star has significant atomic line blanketing not included in the modeling. In Figure 2, we created temperature–density plots that show the actual structure of the disk and that will allow us to see what changes structurally from model to model. In this model, the structure of the innermost region contains out-of-plane material, not unlike the disk wind of Bans & Königl (2012) for Herbig Ae stars (including HD 163296), and for similar reasons—to obtain added NIR emission difficult to achieve with a disk alone, and similar to the one employed by Sitko et al. (2008) for HD 163296.

### 3.2. Model 2: 2001 V Band Drop in Brightness

To achieve the large dip in brightness shown in our data in Figure 3, we added some material to the bipolar cavity. We also lowered the fiducial density at 1.3 au of the envelope while increasing the exponent of the power-law density for the envelope. By doing so, we created the jet-like structure for the star with a disk wind. The material in the jet is simultaneously heated and blocks some of the light in the V band (Ellerbroek et al. 2014) from the observer's line of sight. The heated material creates a large increase in brightness in the 1–8 μm region. Unfortunately, no published NIR photometry exists to compare the model to in this wavelength region. This material also shines down on portions of the disk, which may have been previously shadowed by the inner rim, raising the flux even at much longer wavelengths.

In the temperature–density plot of this epoch, Figure 3, we can see the increase in both temperature and density in the bipolar cavity and the closest region to the star that agrees with our theoretical model.

### 3.3. Model 3: Modeling 2002 Brightening of the NIR Region

In 2002, the BASS data indicated an increase in brightness persisted a year after the major ejection event in 2001. This indicates that, while the dust was no longer in the line of sight to the star, much of it was still warm enough to be detected in the near-to-mid IR (the power-down phase). Our model (Figure 4) reproduces reasonably well both the unusually large IR emission seen in the BASS data, but also the recovery of the brightness in the V band. In the temperature–density plot in Figure 4 supports our model. The cavity still has some material in it, so we can see that everything is still warmer than in 2005 while the widening of the opening angle of the outer cavity surface after 2001 could be because after some time the material spreads out. The temperature around the star and in the innermost region is still warmer than in 2005.

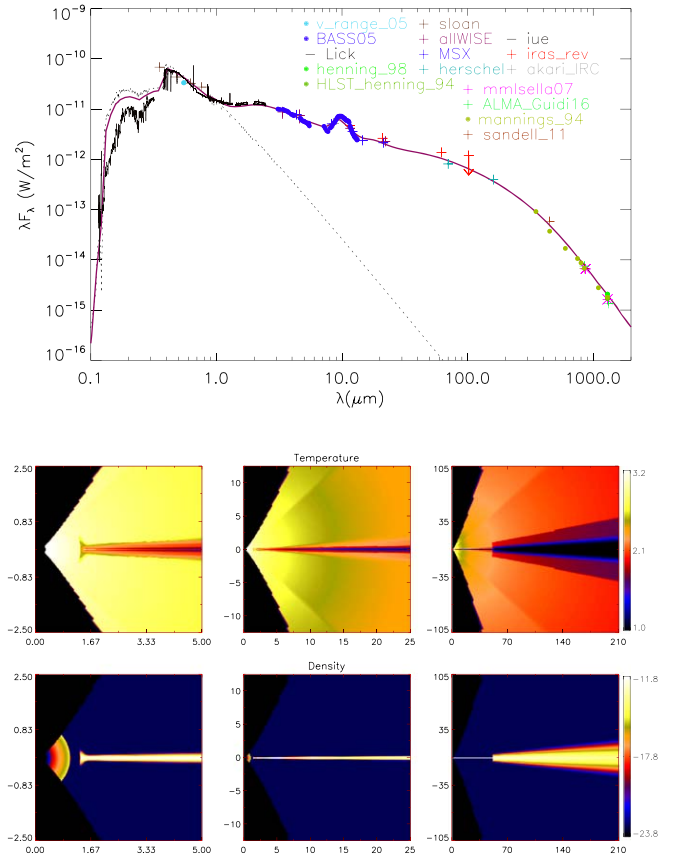

Figure 2. Top panel: SED model of Model 1, the quiescent state of HD 163296. The different data sets and points are color coded and named based on either the instrument that had done the observation or by the author of the paper who published them, and for some both. The magenta line is our model and the dotted line is a model of the star exclusively. Bottom panel: Temperature–density plots of our 2005 model, Model 1. The top three plots show the temperature distribution of $\log_{10} T$ in kelvin over three distances from the star in astronomical units. The bottom three plots show the density distribution of $\log_{10} \rho$ in g cm$^{-3}$ over the same three distances from the star as the temperature plots. From left to right, the empty bipolar cavity (in black), the small puffed up inner rim, the fan, and the start of the small grain disk around 64 au are clearly visible on the plots.

The envelope properties stay the same as in our 2001 model, but the outer cavity surface exponent was lowered while the opening angle of the cavity was increased. The actual values can be found in Table 3.

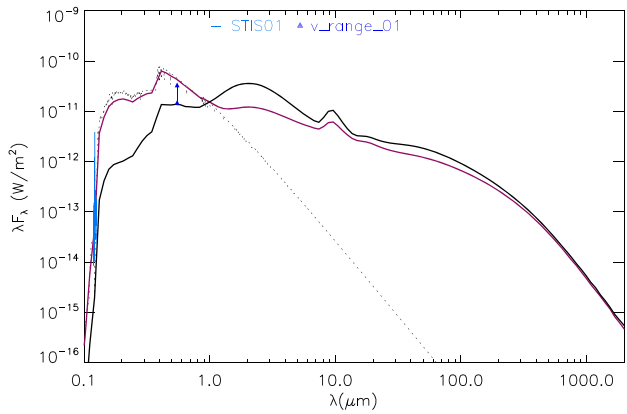

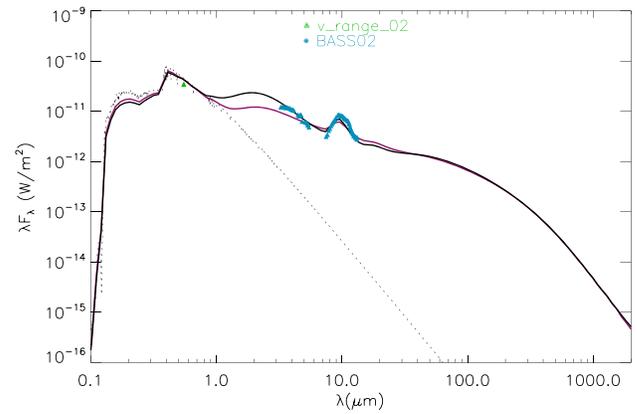

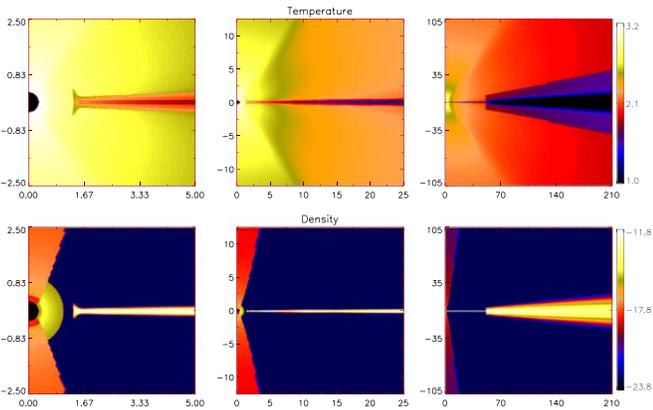

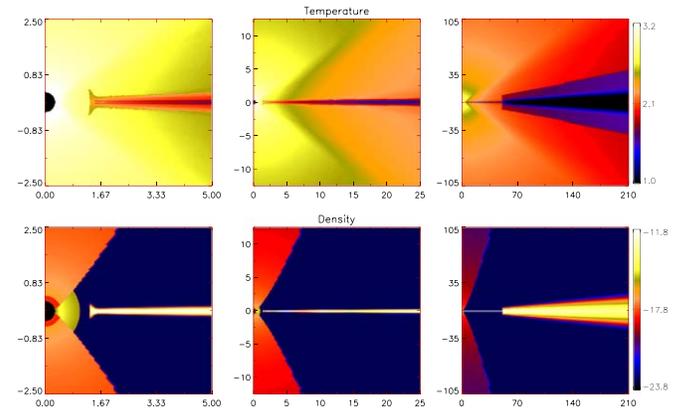

**Figure 3. Top panel:** SED model of Model 2, the 0.8 magnitude drop in brightness in the V band. We show the data obtained in 2001 (range between two dark blue triangles and light blue STIS data) and our model to fit the data (black line). To see the difference from the quiescent model we added the magenta line to this plot. The dotted line is a model of the star exclusively. **Bottom panel:** Temperature–density plots of our 2001 model. The top three plots show the temperature distribution of $\log_{10} T$ in kelvin over three distances from the star in astronomical units. The bottom three plots show the density distribution of $\log_{10} \rho$ in g cm$^{-3}$ over the same three distances from the star as the temperature plots. From left to right, the filled cavity in both sections, the small puffed up inner rim, the fan, and the start of the small grain disk around 64 au are clearly visible on the plots.

**Figure 4. Top panel:** SED model of Model 3, the 2002 state of HD 163296. The plot shows only the data obtained in 2002 on the model (black line) while the quiescent model (magenta line) shows the leftover increase in brightness in 1–8 μm wavelengths. The dotted line is a model of the star exclusively. **Bottom panel:** Temperature–density plots of our 2002 model. The top three plots show the temperature distribution of $\log_{10} T$ in kelvin over three distances from the star in astronomical units. The bottom three plots show the density distribution of $\log_{10} \rho$ in g cm$^{-3}$ over the same three distances from the star as the temperature plots. The widening of the opening angle is of the outer cavity surface, the small puffed up inner rim, the fan, and the start of the small grain disk around 64 au are visible (from left to right).

## 4. Discussion and Summary

In this paper, we examined the star HD 163296 at one point of its evolution. At this stage, the system consists of a star, a disk, and bipolar outflows of material that produce Herbig–Haro objects with measurable proper motions. Ellerbroek et al. (2014) reported that in 2001 a significant drop in brightness of the star at visible wavelengths occurred, and they suggested that this was dust entrained in a disk wind, similar to that used by Bans & Königl (2012) to explain the origin of the NIR emission in Herbig Ae stars. A rotating disk wind for HD 163296 has also been suggested by Pojmanski & Maciejewski (2004), based on ALMA imaging. Data obtained from the Keck Interferometer in 2003 by Tannirkulam et al. (2008), which were sensitive to material along the jet axis, deviated from the disk model that fit the bulk of the available interferometry, and we identify this as the detection of some of the jet material launched in a wind in 2001.

In order to investigate the disk wind structure suggested by Ellerbroek et al. (2014) in a more quantitative way, we have modeled the star+disk system in three different states: its quiescent state, and two epochs where the brightness varied—in 2001 in the visible light and in 2002 in the NIR, respectively. We have presented a jet-like model for the star and the possible causes for the changes in brightness in 2001 and later in 2002, which supports the general picture of material launched around 2000–2001, which occulted much of the light from the star (2001) and increased the thermal emission at NIR wavelengths (2002) before returning to a more quiescent state. If our model is correct, an even greater degree of emission at NIR wavelengths likely occurred at the same time as the occultation event, which was missed by our observations and no other NIR photometry exists to compare the model to in this region. If such an event happened again in the future, it would be beneficial to look for similarities. Based on Figure 5 in the publication of Ellerbroek et al. (2014), we could speculate that the periodicity of the system is around 15 yr but in 2016, our observations did not detect any activity. There could be multiple reasons for this: (i) the event could have lasted only a short amount of time (Sitko et al. 2018), so it is possible that we did not detect it within the allotted time; (ii) the event was first detected with BASS and we were not granted time on BASS in 2016, so we might have missed it because BASS operates in the wavelengths needed for such detection. We will have to depend on future observations, especially in the NIR region, in order to better constrain our models.

The following are the main results of this work:

1. The MCRT code can replicate the quiescent state of the system.
2. The ∼0.8 mag drop in brightness of the V photometric band seen in 2001 by the ASAS (Pojmanski & Maciejewski 2004) can be reproduced by introducing a jet-like structure for the star with a disk wind. The structure is produced by introducing a bipolar cavity filled with material to the code. The material in the jet can be simultaneously heated, and by blocking some of the light in the V band, create the dip that is observed. However, the heated material creates a large increase in brightness in the 1–8 $\mu$m region.
3. The subsequent ∼50% increase in brightness in NIR reported by Sitko et al. (2008) can be reproduced sufficiently by the cavity still having some material that is spreading out while the dust is no longer in the line of sight. The BASS 2002 data agrees with the suggested model and shows that the increase in brightness persisted a year after the major ejection from the jet in 2001.

While our models agree reasonably well with the observations and could be reproduced to study similar events in the future, a more rigorous test can be made with truly contemporaneous visible and NIR observations. Such an investigation could be done with a combination of ground-based visible/IR observations and space telescopes. Coronagraphic imaging of the disk is possible with VLT/SPHERE, Subaru/SCExAO, and in the future with JWST. These could be used to see how disk illumination is affected by jet/wind activity. Monitoring observations with LBTI and VLTI/GRAVITY/MATISSE could be pursued, for more direct measurements of the innermost structure, as was done with the Keck Interferometer.

M.L.S., C.A.G., and W.C.D. are supported by NASA Exoplanet Research Program NNX16AJ75G. M.L.S., J.P.W., E.A.R., and C.A.G. are funded by NNX17AF88G. C.A.G. is additionally supported under NASA Origins of Solar Systems Funding via NNG16PX39P, and J.P.W. and E.A.R. are supported from NSF-AST 1009314 and the NASA Origins of Solar Systems program under NNX13AK17G. This research had made use of the NASA/IPAC Infrared Science Archive, which is operated by the Jet Propulsion Laboratory, California Institute of Technology, under contract with the National Aeronautics and Space Administration. The results reported therein benefited from collaborations and information exchange within NASA's Nexus for Exoplanet System Science (NExSS) research coordination network sponsored by NASA's Science Mission Directorate.

This work has made use of data from the European Space Agency (ESA) mission Gaia (https://www.cosmos.esa.int/gaia), processed by the Gaia Data Processing and Analysis Consortium (DPAC, https://www.cosmos.esa.int/web/gaia/dpac/consortium). Funding for the DPAC has been provided by national institutions, in particular the institutions participating in the Gaia Multilateral Agreement.

*Software:* Spextool (Cushing et al. 2004), HOCHUNK3D (Whitney et al. 2013).

## Appendix A
## Observations

The dates and airmasses of HD 163296 and HD 163336 during the SpeX and IRTF observations are listed in Table 4.

**Table 4**
List of Observations

| UT Date    | Telescope | Instrument | Comments      |
|------------|-----------|------------|---------------|
| 1996.10.14 | IRTF      | BASS       |               |
| 1998.05.12 | IRTF      | BASS       |               |
| 2002.07.18 | Lick 3 m  | NIRIS      |               |
| 2002.07.29 | IRTF      | BASS       |               |
| 2004.05.14 | MLOF      | BASS       | Low Quality   |
| 2004.05.29 | IRTF      | BASS       |               |
| 2004.05.31 | MLOF      | BASS       | Low Quality   |
| 2004.06.01 | MLOF      | BASS       | Low Quality   |
| 2004.06.16 | MLOF      | BASS       |               |
| 2005.07.06 | Lick 3 m  | VNIRIS     |               |
| 2005.07.26 | IRTF      | BASS       |               |
| 2006.04.20 | IRTF      | SpeX       |               |
| 2006.05.18 | IRTF      | BASS       | Low Quality   |
| 2006.05.19 | IRTF      | BASS       |               |
| 2006.08.23 | IRTF      | SpeX       |               |
| 2007.05.01 | IRTF      | SpeX       |               |
| 2007.07.08 | IRTF      | SpeX       | SXD only      |
| 2007.08.17 | IRTF      | SpeX       |               |
| 2007.07.17 | IRTF      | BASS       |               |
| 2007.08.21 | IRTF      | SpeX       |               |
| 2008.05.22 | IRTF      | SpeX       |               |
| 2008.05.22 | AEOS      | BASS       |               |
| 2008.09.03 | IRTF      | BASS       |               |
| 2008.09.05 | IRTF      | SpeX       |               |
| 2009.07.08 | IRTF      | SpeX       |               |
| 2009.07.14 | IRTF      | BASS       |               |
| 2009.07.16 | IRTF      | BASS       |               |
| 2011.03.22 | IRTF      | SpeX       |               |
| 2011.03.23 | IRTF      | SpeX       |               |
| 2011.04.19 | IRTF      | SpeX       |               |
| 2011.04.28 | IRTF      | SpeX       |               |
| 2011.07.30 | IRTF      | BASS       |               |
| 2011.07.31 | IRTF      | SpeX       |               |
| 2011.08.01 | IRTF      | BASS       | Poor 3–4 $\mu$m |
| 2013.06.17 | IRTF      | SpeX       |               |
| 2013.07.16 | IRTF      | SpeX       | SXD only      |
| 2013.07.17 | IRTF      | SpeX       | LXD only      |
| 2016.04.04 | IRTF      | SpeX       |               |
| 2016.05.04 | IRTF      | SpeX       |               |
| 2016.05.12 | IRTF      | SpeX       |               |

## Appendix B
## Dust Extinction of the Outflow

While we are unaware of multiwavelength visible-wavelength photometry obtained on HD 163296 during one of its major extinction events, we did obtain BVRI photometry during 2012 from which we can derive the extinction properties of the dust. In Figure 5 we show the brightness of HD 163296 between 2011 March 27 and 2012 November 5 (UT) in the BVRI filters obtained with the Bright Star Monitor telescopes of the AAVSO.[13] By comparing the brightness on the one faintest state with the median values in each filter, we produced an extinction curve of the dust, assuming that the drop was due to a temporary increase in dust obscuration. The result, shown in Figure 5, is compared to different reddening curves, computed using different values of $R_V$ and the extinction curves derived by Cardelli et al. (1989). It is evident that standard ISM extinction with $R_V = 3.1$ mag is excluded. Instead, larger grains, producing values of $R_V \geqslant 5$ are indicated. However, none of these curves matched the observed extinction, suggesting that the dust did not follow the standard extinction prescription of Cardelli et al. (1989).

---

[13] https://www.aavso.org/bright-star-monitor-stations

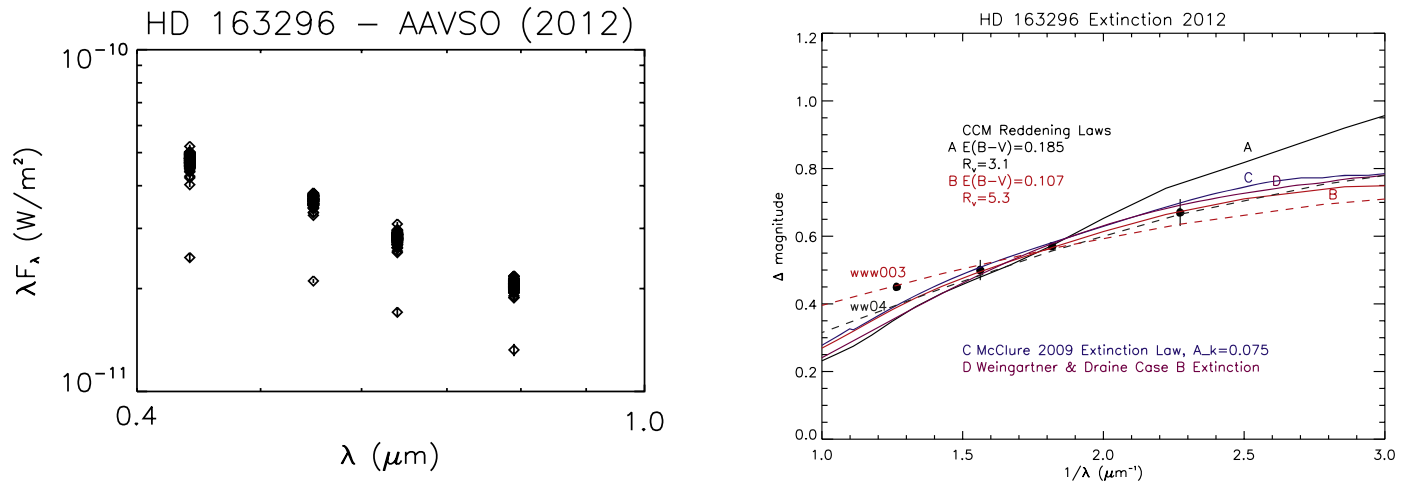

**Figure 5. Left:** BVRI brightness of HD 163296 between 2011 March 27 and 2012 November 5 (UT). **Right:** The extinction of the one day event where the brightness dropped by roughly 1/2 magnitude. A number of different extinction laws are shown. The standard interstellar extinction of Cardelli et al. (1989) with $R_V = 3.1$ mag is shown, along with another with $R_V = 5.3$ mag for larger grains. Also shown are the large-grain models of Weingartner & Draine (2001) and McClure (2009).

For the disk wind component, we used the standard ISM extinction based on Kim et al. (1994), as warm, small grains are required to maintain the strength of the 10 $\mu$m silicate band. As our model is rotationally symmetric, the silicate emission band includes all of the dust in this structure, while the existence of small clumps of bigger grains cannot be excluded. If the main extinction events are produced by such clumps, then the drop in extinction at shorter wavelengths will be considerably smaller than those produced in our model.

## Appendix C
## Mass and Age of HD 163296

Using the V magnitude of HD 163296, the distance from the second Gaia data release, the effective temperature of 9400 K from Guimarães et al. (2006), and the bolometric correction and effective temperature for PMS stars from Pecaut & Mamajek (2013) we derived a luminosity of $20.4 \pm 0.5 \, L_\odot$. Figure 6 shows the location of HD 163296 relative to the PMS evolution tracks and isochrones of the nongray (lines, discontinuities; Tognelli et al. 2011), for $z = 0.02$, helium mass fraction of 0.27, convective parameter $\alpha = 1.68$, and a deuterium abundance of $2 \times 10^{-5}$. The online versions of these use the abundances of Asplund et al. (2005) but are nearly indistinguishable from those based on the revised abundances of Asplund et al. (2009) as seen in the figures in Tognelli et al. (2011). From these we derived a mass of $2.14 \pm 0.06 \, M_\odot$ and an age of $9.0 \pm 0.5$ Myr. The values for the mass and age differ slightly from those derived by Vioque et al. (2018) ($1.83 \pm 0.09 \, M_\odot$ and $7.60 \pm 1.1$ Myr), who adopted an effective temperature of 9200 K and the PMS tracks and isochrones of Bressan et al. (2012) and Marigo et al. (2017).

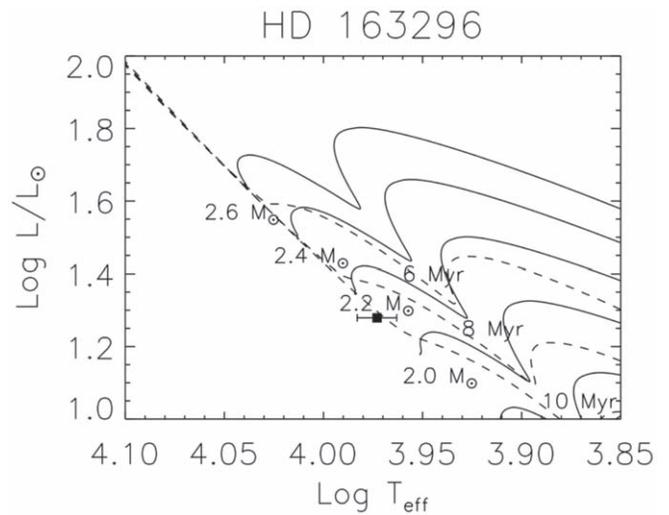

**Figure 6.** The mass and age of HD 163296. Using the parallax from the Gaia data release 2, the pre-main sequence star parameters of Pecaut & Mamajek (2013), and the evolutionary tracks and isochrones of Tognelli et al. (2011), we derive a mass of $2.14 \pm 0.06 \, M_\odot$ and an age of $9.0 \pm 0.5$ Myr. Previous values of the mass and age placed the star well above the region of the main sequence, but the Gaia parallax places the star at 101.5 pc, rather than the previous value of ~120 pc, thus lowering the luminosity of the star compared to past values.

## Appendix D
## Derivation of Mass Accretion Rates

The mass accretion rates for each epoch of SpeX data were derived using the same procedures as those described by Sitko et al. (2012). The absolute flux-calibrated spectra were modeled as a combination of a stellar photosphere matched to the spectral type of HD 163296, plus a thermal component due to the hottest dust component in the system. For the stellar template spectrum, we used SAO 206463, a pre-main sequence A0V star exhibiting neither significant gas accretion nor thermal dust emission, and for which we had comparable SpeX data. To this, we added a modified blackbody to approximate the thermal emission of the innermost dust.

Because the models were not able to produce a pseudo-continuum that matched the data in every spectral order at every line to be extracted, the model was adjusted locally using a vertical scaling until the $\chi^2$ of the difference in the continuum nearby (but outside) the line was minimized. For the majority of the lines on all nights, these corrections to the scaling were less than 2% of the initial model continuum level.

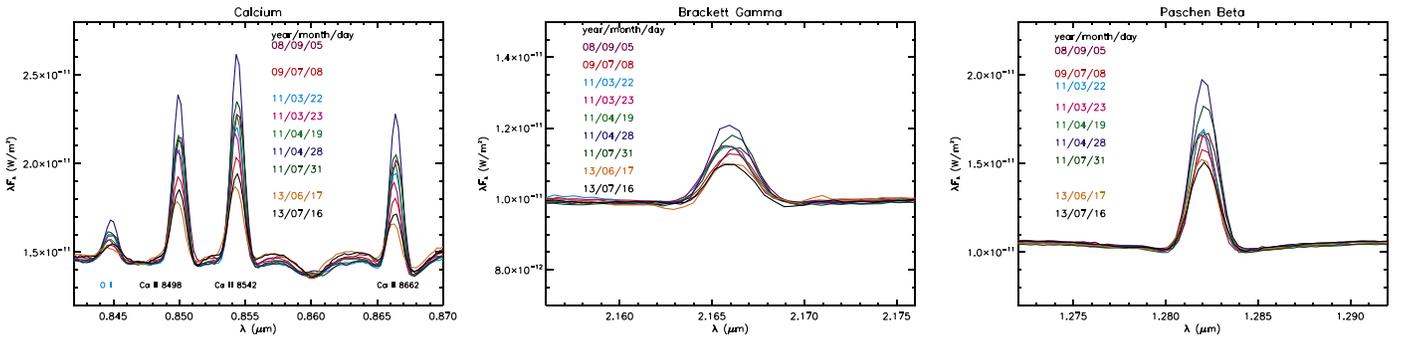

**Figure 7.** From left to right: extraction lines of Calcium II, Brackett $\gamma$, and Paschen $\beta$ over multiple years.

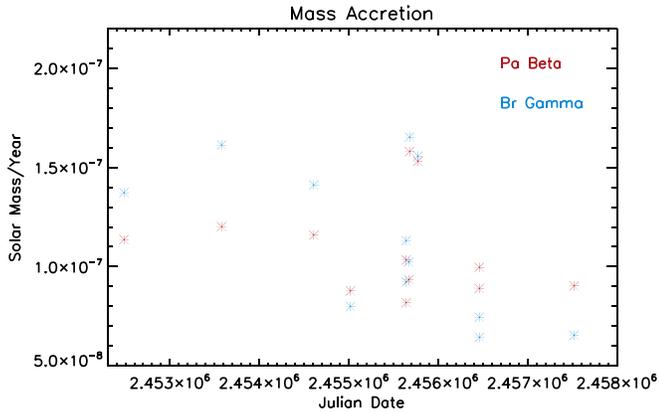

**Figure 8.** Paschen $\beta$ (red) vs. Brackett $\gamma$ (blue) mass accretion over time. Mass and radius from Section 3 and calibration from Fairlamb et al. (2017) were used for accretion rates calculations.

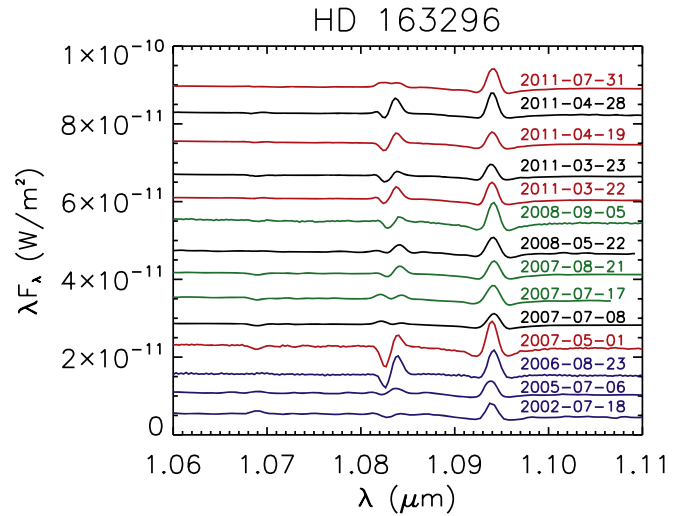

**Figure 9.** A succession of epochs of SpeX spectra, stacked for easy comparison. The He I line is highly variable, and while usually exhibiting some form of P Cygni profile indicative of gas outflow, a number of epochs have a double-humped emission profile, suggesting a more complicated flow geometry.

The line strengths of the Pa$\beta$ and Br$\gamma$ lines (Figure 7) were then extracted by subtraction of the adjusted model from the flux-calibrated data, and the net line flux calculated by integrating over the flux-calibrated continuum-subtracted line profile. These were converted to line luminosities using a distance of 101.5 pc. The line luminosities were transformed to mass accretion luminosities ($L_{acc}$) using the calibrations of Fairlamb et al. (2017) for Pa$\beta$ and Br$\gamma$, respectively. These were then converted to mass accretion rates using $\dot{M} = L_{acc} R_*/GM_*$, with adopted values of $R_* = 1.71 R_\odot$, and the stellar mass $M_* = 2.14 M_\odot$. While there are many different calibrations of the mass accretion rates based on these line, we selected that of Fairlamb et al. (2017), as it was derived specifically for Herbig AeBe stars, and has among the smallest calibration uncertainties in the literature. The mass accretion rates were time dependent, being $(6.4–16.5) \times 10^{-8} M_\odot$ yr$^{-1}$ and $(8.8–15.8) \times 10^{-8} M_\odot$ yr$^{-1}$ derived from Br$\gamma$ and Pa$\beta$ lines, respectively. In Figure 8, the calculated mass accretion rates of Paschen $\beta$ versus Brackett $\gamma$ are shown over time.

We note that the changes in the strength of the Ca II IR triplet lines (Figure 7) seem to be slightly greater than those of the hydrogen lines. This suggests that some additional variability might be due to a magnetically active chromosphere. While A stars are not supposed to be magnetically active, Grady et al. (2010) found that the accretion onto the Herbig Ae star MWC 480 was primarily confined to mid latitudes, suggesting possible magnetic collimation.

In Figure 9, we show a subset of existing SpeX spectra of HD 163296 between 2002 and 2011. The He I line, which is a good tracer of gas motion, is highly variable in strength and shape. Usually, it exhibits the standard P Cygni profile indicative of an outflowing wind. However, on numerous occasions, it exhibits a "double-humped" emission profile. This double-humped profile is consistent with the additional presence of an inverse P Cygni profile, indicative of infalling gas. These events could well indicate epochs where the observer is actually looking down an accretion column. See Figure 10 for the double-humped emission profile from 2011.

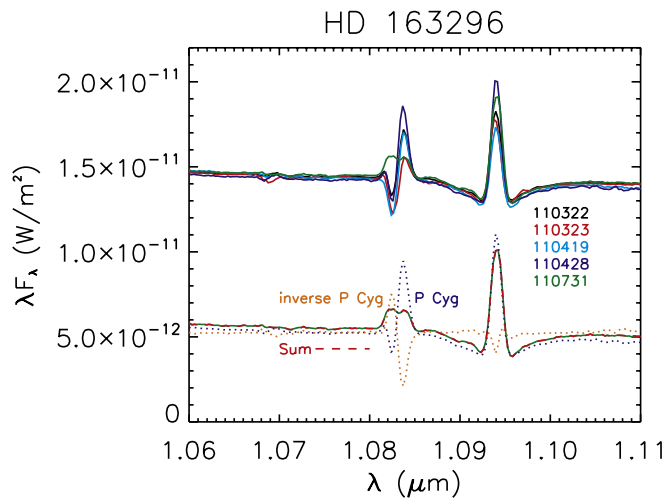

**Figure 10.** SpeX data from 2011, showing one night, 2011 July 31 (110731 in plot), where the double-humped emission profile was present. A sum of a P Cygni profile and inverse P Cygni profile was constructed using the spectrum from 2011 April 19 (110419) and an inverted spectrum from 2011 April 28 (110428), both scaled appropriately to achieve a reasonable agreement with the spectrum of 2011 July 31 in the region of the He I line. The presence of an inverse P Cygni component at some epochs suggests that we were looking down an accretion column, possibly guided by a weak magnetic field, as was suggested for MWC 480 by Grady et al. (2010). The frequency with which the inflow is observed provides a rough idea of the fraction of the stellar surface that was involved.

## ORCID iDs

Monika Pikhartova ● https://orcid.org/0000-0002-6767-6377
Zachary C. Long ● https://orcid.org/0000-0002-2048-5741
Rachel B. Fernandes ● https://orcid.org/0000-0002-3853-7327
Michael L. Sitko ● https://orcid.org/0000-0003-1799-1755
Carol A. Grady ● https://orcid.org/0000-0001-5440-1879
John P. Wisniewski ● https://orcid.org/0000-0001-9209-1808
Evan A. Rich ● https://orcid.org/0000-0002-1779-8181